# Imaging Light-Induced Migration of Dislocations in Halide Perovskites with 3D Nanoscale Strain Mapping

Kieran W. P. Orr[1,2], Jiecheng Diao[3,4], Muhammad Naufal Lintangpradipto[5], Darren J. Batey[6], Affan N. Iqbal[1,2], Simon Kahmann[1,2], Kyle Frohna[1], Milos Dubajic[2], Szymon J. Zelewski[1,2], Alice E. Dearle[1,2,7], Thomas A. Selby[2], Peng Li[6], Tiarnan A. S. Doherty[1,2,8], Stephan Hofmann[7], Osman M. Bakr[5], Ian K. Robinson[3,9], Samuel D. Stranks*[1,2]

**ABSTRACT:**

In recent years, halide perovskite materials have been used to make high performance solar cell and light-emitting devices. However, material defects still limit device performance and stability. Here, we use synchrotron-based Bragg Coherent Diffraction Imaging to visualise nanoscale strain fields, such as those local to defects, in halide perovskite microcrystals. We find significant strain heterogeneity within MAPbBr$_3$ (MA = CH$_3$NH$_3^+$) crystals in spite of their high optoelectronic quality, and identify both ⟨100⟩ and ⟨110⟩ edge dislocations through analysis of their local strain fields. By imaging these defects and strain fields *in situ* under continuous illumination, we uncover dramatic light-induced dislocation migration across hundreds of nanometres. Further, by selectively studying crystals that are damaged by the X-ray beam, we correlate large dislocation densities and increased nanoscale strains with material degradation and substantially altered optoelectronic properties assessed using photoluminescence microscopy measurements. Our results demonstrate the dynamic nature of extended defects and strain in halide perovskites and their direct impact on device performance and operational stability.

[1]*Department of Physics, Cavendish Laboratory, University of Cambridge, JJ Thomson Avenue, Cambridge, CB3 0HE, UK*
[2]*Department of Chemical Engineering and Biotechnology, University of Cambridge, Philippa Fawcett Drive, Cambridge, CB3 0AS, UK*
[3]*London Centre for Nanotechnology, University College London, London WC1E 6BT, UK*
[4]*Center for Transformative Science, ShanghaiTech University, Shanghai 201210, China*
[5]*Physical Science and Engineering (PSE) Division, King Abdullah University of Science and Technology (KAUST) Thuwal 23955-6900, Kingdom of Saudi Arabia*
[6]*Diamond Light Source, Harwell Science and Innovation Campus, Fermi Ave, Didcot OX11 0DE, UK*
[7]*Department of Engineering, University of Cambridge, UK*
[8]*Department of Materials Science & Metallurgy, University of Cambridge, 27 Charles Babbage Road, Cambridge, CB3 0FS, United Kingdom*
[9]*Condensed Matter Physics and Materials Science Department, Brookhaven National Lab, Upton, New York 11793, USA*

*sds65@cam.ac.uk

This work was carried out while J.D. was affiliated with [3]. J.D. is now affiliated with [4].

**INTRODUCTION:**

Halide perovskites are promising materials for highly efficient optoelectronic devices. In under a decade, the power conversion efficiency of halide perovskite-based single junction solar cells has increased from 14.1% (2013)[1] to 25.7% (2021)[2], and efficiencies of 32.5% have been reported for perovskite/Si tandem devices[2]. This tremendous rise in device efficiency has, in large part, been achieved through empirical optimisation of device processing, and the community's fundamental understanding of halide perovskite materials, especially the impact of nanoscale structure, lags behind device performance enhancement.

One aspect of material structure that is poorly understood in halide perovskites (compared to traditional semiconductors) is strain, which has been proposed to affect carrier lifetime[3], bandgap[4], Urbach energy[5], ion migration[6], material stability[6,7], as well as overall device efficiency[8]. In the halide perovskite field, strain is most commonly characterised by assessing Bragg peak shifts (tensile and compressive strain[3–5]) or Bragg peak broadening (microstrain[9]) which evaluate the linear expansion of the material and $d$-spacing disorder, respectively. In general, these are bulk techniques, allowing determination of the average strain state of a material and, even with synchrotron nano-probe facilities, it is difficult to achieve a spatial resolution (spot size) below *ca.* 10 nm[10]. A full description of strain in materials (structural deformation due to applied stress[11]) is derived from atomic displacement vectors, $\mathbf{u}(\mathbf{r})$ (where $\mathbf{r}$ is a real space position vector), which describe the amount by which atoms are displaced from their expected positions according to the underlying lattice. Interrogation of atomic displacements allows one to identify point defects and dislocations from their characteristic local strain fields, providing unparalleled insight into the internal structure of materials. $\langle 100 \rangle$ edge dislocations have been reported in the halide perovskite $FAPbI_3$[12] (FA = $CH(NH_2)_2^+$) and $\frac{1}{2}\langle 110 \rangle$ screw dislocations have been identified in $CsPbBr_3$[13], both using atomic-resolution scanning transmission electron microscopy (STEM). However, STEM and related electron microscopy techniques generally require use of thin samples only (<< 10 nm), rendering them of limited use for imaging buried dislocations in crystals relevant for optoelectronic devices. Electron tomography, while able to image dislocations in semiconductors in 3D[14], uses beam doses that are prohibitively high for halide perovskites.

Bragg coherent diffraction imaging (BCDI) measurements can provide information on atomic displacement vectors with nanometre resolution in thicker samples, such as those that may be used in optoelectronic devices. The technique has been employed to reveal the 3D atomic displacement fields within ZnO nanorods[15], to track the evolution of ferroelastic domain walls in barium titanate[16], and to identify twin domains in $CsPbBr_3$ nanoparticles[17]. Further, the approach has been used to monitor the growth of crystalline grains during the annealing process of $In_2O_3$:Zr thin films[18], during calcite crystal solution and dissolution[19], and to track dislocations dynamics in the $LiNi_{0.5}Mn_{1.5}O_4$ battery electrode material during charging and discharging[20].

Here, we develop an *in situ* BCDI approach to visualise the surprisingly rich strain fields within high quality single microcrystals of halide perovskites and monitor their evolution under continuous solar illumination, a first in the BCDI and halide perovskite fields. By tracking the strain fields in $MAPbBr_3$, we identify $\langle 100 \rangle$ and $\langle 110 \rangle$ edge dislocations and observe their surprisingly extensive migration through the crystal structure under illumination. By considering crystals that become damaged under X-ray exposure, we also find dislocation formation to be associated with degradation of the halide perovskite material and changes in its optoelectronic properties. These findings give mechanistic insight into the structural evolution of halide perovskites under operating conditions, and identify the important role extended defects and nanoscale strain play in device performance and operational stability.

**RESULTS AND DISCUSSION:**

**Imaging nanoscale strain with Bragg coherent diffraction imaging:**

BCDI is a lensless imaging technique that involves illuminating a sample with a coherent beam of X-rays from a synchrotron source. When the diffracting object is the same size or smaller than the lateral coherence length of the X-ray beam, a coherent X-ray diffraction pattern can be recorded that includes extra interference fringes seen around each of the Bragg peaks. An example coherent diffraction pattern is shown on the detector in Fig. 1a. From a set of these patterns, collected at different sample rocking angles, a three-dimensional electron density function $\rho(\mathbf{r})$ can be reconstructed using an iterative phase retrieval algorithm, details of which are given in the Methods and elsewhere[21–23]. In general, $\rho(\mathbf{r})$ is complex-valued, with the modulus proportional to the crystal's electron density and the argument proportional to the size of the atomic displacement along the direction of the X-ray scattering vector, $\mathbf{Q}$[24,25]. To a first approximation, the size and shape of the diffracting crystal can be extracted from the interference fringe spacing around the outside of the Bragg peak, and the strain information is contained in the intensity asymmetry near the centre of the peak[24]. Fig. 1b shows a 3D rendering of the reconstructed electron density of an example microcrystal of MAPbBr$_3$, with scanning electron micrographs (SEM) of representative microcrystals in Fig. 1c & d (microcrystal films and isolated microcrystals, respectively). Henceforth, such electron density renderings will be referred to simply as "reconstructions". Two perpendicular slices taken through the reconstruction are also shown in Fig. 1b and the colour scale indicates the size of the atomic displacement along the scattering direction.

By taking the spatial derivative of the displacement field of the reconstruction shown in Fig. 1b with respect to the direction of the scattering vector, we obtain values for the local nanoscale tensile (positive) and compressive (negative) strain between voxels of the reconstruction. These values correspond to one of the local diagonal elements of the microscopic strain tensor. Fig. 1e shows a histogram of these local strain values for the crystal in Fig. 1b, with those of magnitude greater than 1% highlighted in orange. For this crystal, the fraction of the crystal volume with local strain exceeding 1%, $f$, is 13.2%, and the root mean squared value of local strain, $\epsilon_{\mathrm{rms}}$, is 0.7%. These values are relatively low when compared to those for other (undamaged) crystals in this study, for which we find $5\% < f < 30\%$ and $0.5\% < \epsilon_{\mathrm{rms}} < 4.5\%$ (see Supplementary Table 1). Such strain values are remarkably high since halide perovskites prepared in this manner are used in high-performing devices[26]. By contrast, devices based on Si[27] and Cu(In,Ga)Se$_2$[28] suffer significant performance losses after strain exceeds *ca.* 1%. Such an observation links closely to the apparent defect-tolerance in halide perovskites, where such large strains are not catastrophic for optoelectronic performance.

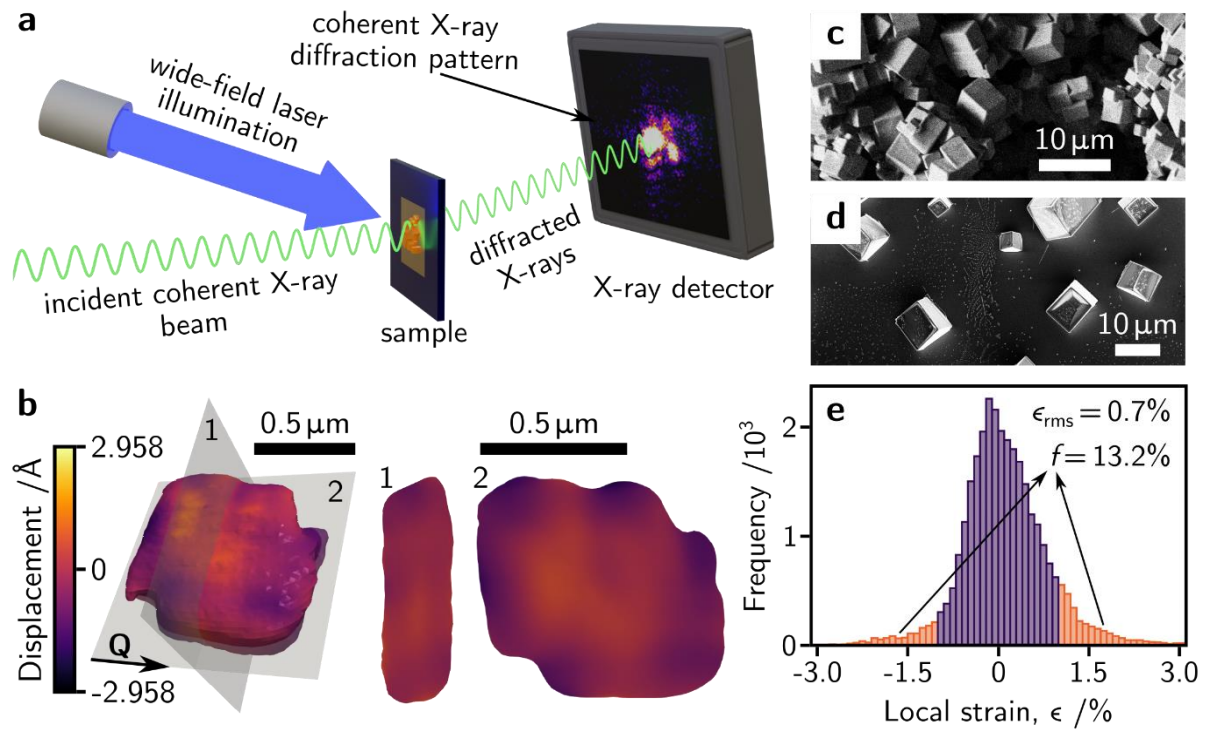

**Fig. 1: BCDI enables reconstruction of strain fields in halide perovskite microcrystals: a** Schematic of the BCDI measurement geometry. A coherent beam of X-rays is incident from the left and is diffracted by crystals on the sample towards a detector that records a coherent X-ray diffraction pattern. A laser mounted on the beamline provides wide-field visible light illumination with a wavelength of 405 nm. **b** Example electron density reconstruction of a MAPbBr$_3$ microcrystal (left), with slices through the 3D volume indicated by the numbered grey planes (middle and right). The colour scale indicates the size of the atomic displacement along the X-ray scattering vector direction. Scanning electron microscopy images of representative crystals of **c** a microcrystal film (continuous films made up of many touching microcrystals) and **d** isolated MAPbBr$_3$ microcrystals. **e** Histogram of local nanoscale tensile (positive) and compressive (negative) strain for the crystal shown in panel **b**. Local strain values exceeding a magnitude of 1% are highlighted in orange. The root mean squared local strain, $\epsilon_\mathrm{rms}$, and fraction of the crystal volume with local strain exceeding 1%, $f$, are quoted.

**Characterising dislocations in halide perovskites:**

Dislocations can be identified in reconstructions from a core of low electron density surrounded by a characteristic displacement field[16]. Fig. 2a shows a reconstruction of a MAPbBr$_3$ microcrystal rendered in grey and made partially transparent to facilitate identification of a dislocation, which is highlighted with a black line. An infinitesimally thin slice through this reconstruction, perpendicular to the dislocation, is coloured according to the atomic displacements and is shown in the reconstruction in Fig. 2a, and face-on in Fig. 2b.

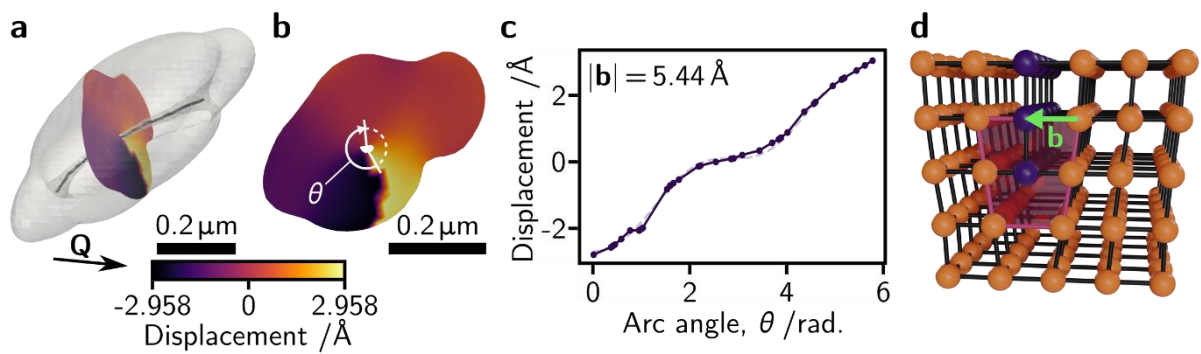

**Fig. 2: Visualising buried edge dislocations in a MAPbBr$_3$ crystal: a** Electron density reconstruction shown partially transparent in grey, with a ⟨100⟩ edge dislocation indicated by the black line. A slice through the reconstruction is shown coloured according to the size of the atomic displacement along the direction of the scattering vector. **b** The slice from **a** shown face-on. **c** Points and solid line: The atomic displacements as a function of arc angle (indicated by the white arrow in **b**). Dashed line: fit to data of the function for atomic displacement, $u$. **d** Schematic representation of an edge dislocation. The atoms in purple belong to an extra atomic plane in the top three rows of the structure. The region in pink highlights the highly strained region in the vicinity of the dislocation and the green arrow is the Burgers vector, **b**.

In Fig. 2c, we show the displacement values as a function of arc angle, $\theta$, here defined as zero at minimum atomic displacement (travelling in a circle around the dislocation core shown by the white arrow in Fig. 2b). The periodic modulation of the displacement superimposed on a linear trend indicates that the dislocation present is an edge dislocation (as opposed to a screw dislocation)[29], which is shown schematically in Fig. 2d. The Burgers vector, **b**, describes the lattice distortion caused by a dislocation[29] and is represented by the green arrow in Fig. 2d. To determine the magnitude of the Burgers vector, |**b**|, we fit the displacement *vs.* arc angle data shown in Fig. 2c according to the function[16,20,30]

$$u = \frac{|\mathbf{b}|}{2\pi}\left(\theta + \frac{\sin(2\theta)-\cos(2\theta)}{4(1-\nu)} + \frac{(2\nu-1)}{2(1-\nu)}\log(r)\right),$$

where $u$ is the size of the atomic displacement (*i.e.* $u = \mathbf{u}(\mathbf{r}) \cdot \mathbf{Q}$), $\nu$ is the Poisson's ratio of the material (taken to be 0.29[31]), and $r$ is the radius of the circle from which we extract the data (see Methods for more details of calculation). For this dislocation, we find $|\mathbf{b}| = 5.44$ Å, which is close to the lattice parameter, $d_{100} = 5.92$ Å (an average of literature values[32–34]). Therefore, we conclude that this dislocation is a ⟨100⟩ edge dislocation. Other possible edge dislocations would involve insertion of lattice planes with smaller interplanar spacings, for example a 110 dislocation with $|\mathbf{b}| = d_{110} = 4.19$ Å, or a 111 dislocaiton with $|\mathbf{b}| = d_{111} = 3.42$ Å, which are in significantly worse agreement with a Burgers vector magnitude of 5.44 Å.

We note that the reconstruction shown in Fig. 2b is not as well-faceted as that shown in Fig. 1b, and the reconstructions appear less well-faceted than expected from the crystal shapes in the SEM images (Fig. 1c & d). Regions of the crystal that are off Bragg condition will not diffract any X-rays to the detector, and so any void regions are likely due to parts of the crystal being twisted into a slightly different orientation, or that are amorphous. Such twisting of the underlying lattice has been reported for halide perovskites; for example, intra-grain orientational heterogeneity has been confirmed from electron backscatter diffraction in MAPbI$_3$[35] (see Supplementary Note 1 and Supplementary Fig. 1). Where reconstructions from multiple scans are shown (for example in Fig. 3), unless stated otherwise, the diffracted intensity remains constant for each scan, indicating that these missing volumes are not caused by beam damage.

**Light-induced dislocation migration:**

To understand the nanoscale structural behaviour under device-like conditions, we now consider the evolution of the strain fields in the halide perovskite crystals under continuous visible illumination. In Fig. 3a we show a reconstruction of a crystal with a bright, structured Bragg peak that was radiation-stable (see Supplementary Note 2 and Supplementary Fig. 2-4). Having found such a crystal, we illuminate the sample with a 405 nm laser at an intensity of *ca.* 1 sun intensity (in terms of photo-generated charge carriers) and simultaneously acquire further BCDI measurements at different time snapshots. Fig. 3b-e show reconstructions of this crystal under continuous illumination with dislocations highlighted in black; each of these dislocations also have Burgers vector magnitudes $\simeq d_{100} = 5.92$Å (Supplementary Fig. 8–12). Under continued illumination, the dislocations show a striking increase in mobility moving hundreds of nanometres through the crystal in minutes compared to when successive measurements are taken in the dark (Supplementary Fig. 2). Thus, we are able to image for the first time the increased migration of such nanoscale extended defects in halide perovskites under visible light illumination. Our result is also consistent with reports showing that ions can migrate more easily under illumination in these materials[36–40]. These results, tracking the migration of buried dislocations in high performance halide perovskites, show their internal structure to be remarkably dynamic under operational conditions.

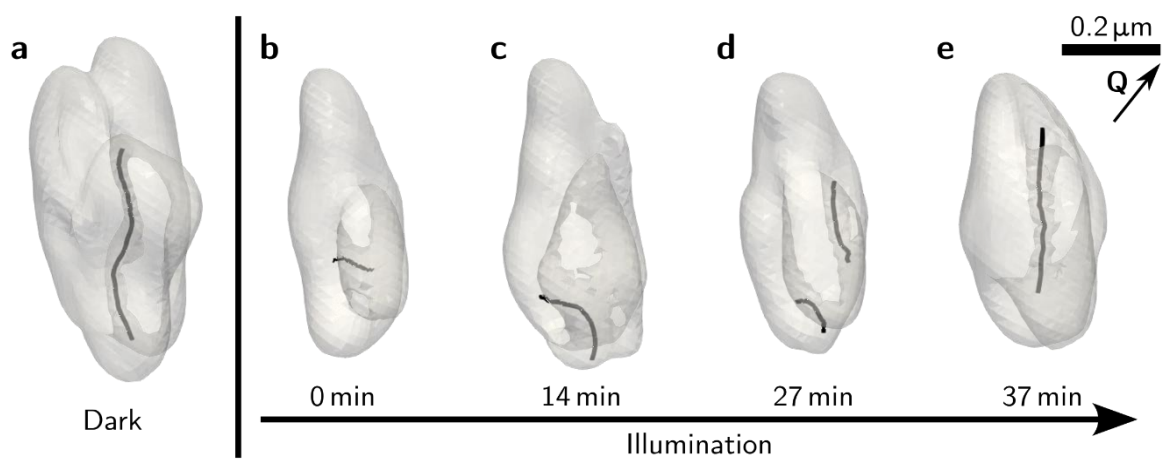

**Fig. 3: Tracking dislocation migration in a MAPbBr$_3$ microcrystal *in situ* under illumination:** Electron density reconstructions of a MAPbBr$_3$ microcrystal from successive BCDI scans. **a** Reconstruction of the crystal in dark conditions. **b**, **c**, **d**, and **e**: Reconstructions of the crystal after continuous illumination with a 405-nm laser at different times. Times given are relative to the start of the BCDI scan giving the reconstruction in **b**. There were 105 minutes of illumination between **a** and **b** for re-optimisation of scan parameters after starting illumination. Dislocations are shown in black and appear very mobile under illumination. The scale bar and scattering vector apply to all reconstructions.

As the dislocations migrate through the crystal under continued illumination, the dislocations curve and take on screw-like character as well as edge-like character, while maintaining a Burgers vector $\simeq$ 5.92 Å. This observation is also reflected in the more linear appearance of displacement *vs.* arc angle plots for some of these dislocations (Supplementary Fig. 8-12). The dislocations also move in three dimensions, *i.e.* not just within a 2D glide plane for a perfect edge dislocation, providing further evidence for the dislocations' mixed character as they evolve and/or evidence of a high density of point defects which may allow the dislocations to climb perpendicular to their glide planes. We note that the volume of the reconstructions shown in Fig. 3 also decreases upon illumination (Supplementary Fig. 13). After BCDI measurements, the crystals are still the same shape (well-faceted) when viewed under PL microscopes with no obvious decrease in size (Supplementary Fig. 14), and therefore we attribute the voids of electron density to regions of the crystals that no longer satisfy the same Bragg condition[41], possibly due to some photo-induced change in orientation of these regions of the crystal. Illumination may also cause point defect formation (as it does in the related MAPbI$_3$ perovskite[42]) which will reduce the structure factor of the crystal, resulting in a smaller reconstructed volume.

**Optoelectronic changes upon dislocation formation:**

In order to understand links between the dislocations, performance and degradation in halide perovskites, we now specifically consider crystals that suffer beam damage during measurement. For these crystals, measurements were carried out using a different, more damaging beam energy of 9.7 keV (as opposed to 11.8 keV for results presented above), and successive scans exhibited lower diffracted intensity indicating a smaller crystal volume on the Bragg condition and/or increasing disorder or point defects in the sample. Reconstructions of such a crystal are shown in Fig. 4a-d. Upon the second exposure to X-rays, the crystal becomes less well-faceted and more strained, as can be seen from the larger and more dramatic changes in colour (atomic displacement). A particularly striking structural change is the formation of a network of many dislocations upon degradation. The majority of these dislocations have Burgers vector magnitudes $\simeq d_{100} = 5.92$ Å, as before (Supplementary Fig. 15-22), but we also identify a dislocation with a Burgers vector of magnitude 4.31 Å $\simeq d_{110}$ which connects two ⟨100⟩ dislocations and is highlighted in the inset of Fig. 4d. We show another example of a crystal that suffers beam damage causing dislocation formation in Supplementary Fig. 23 and Supplementary Fig. 24.

To understand the effects of the increased nanoscale strain and dislocation density in these crystals, we conducted wide field hyperspectral photoluminescence (PL) microscopy and time-resolved confocal PL microscopy measurements on the same sample regions before and after exposure to X-rays. The crystals studied that suffered beam damaged existed in a 20 μm × 20 μm region that was repeatedly exposed to the X-ray beam during the synchrotron experiment and that overlaps with one of these pre-mapped areas (Fig. 4e inset). By summing the PL from this region of the sample before and after the X-ray measurements we find that the damaged crystals have a peak PL emission at 531 nm compared to a value of 541 nm for the pristine crystals (Fig. 4e). Further, we see an increase in PL lifetime for damaged crystals, with the time taken to fall to 1/e of the initial intensity being 0.91 ns for pristine crystals and 2.0 ns for damaged crystals (Fig. 4f) using excitation fluences of 63 nJcm$^{-2}$ and 157 nJcm$^{-2}$ respectively. Such optoelectronic changes could be caused by the formation of nano-sized confined crystallites on the surface of the crystals, or degradation-induced passivation, for example through the formation of PbBr$_2$ (beam-induced PbBr$_2$ formation has been reported for the related (FA$_{0.79}$MA$_{0.16}$Cs$_{0.05}$)Pb(I$_{0.83}$Br$_{0.17}$)$_3$ system)[43]. The structural changes characterised here – dislocation formation, increased strain, significant increase in both $f$ (13.2% to 48.0%) and $\epsilon_{\mathrm{rms}}$ (1.0% to 5.1%), reduction in crystallite volume – clearly correlate with changes in optoelectronic properties. These results have significant implications for the halide perovskites used in devices for photo-detection[44] and direct X-ray detection[45] because dislocations may contribute to non-radiative recombination, afterglow/fall time, responsivity and parasitic absorption[44,46].

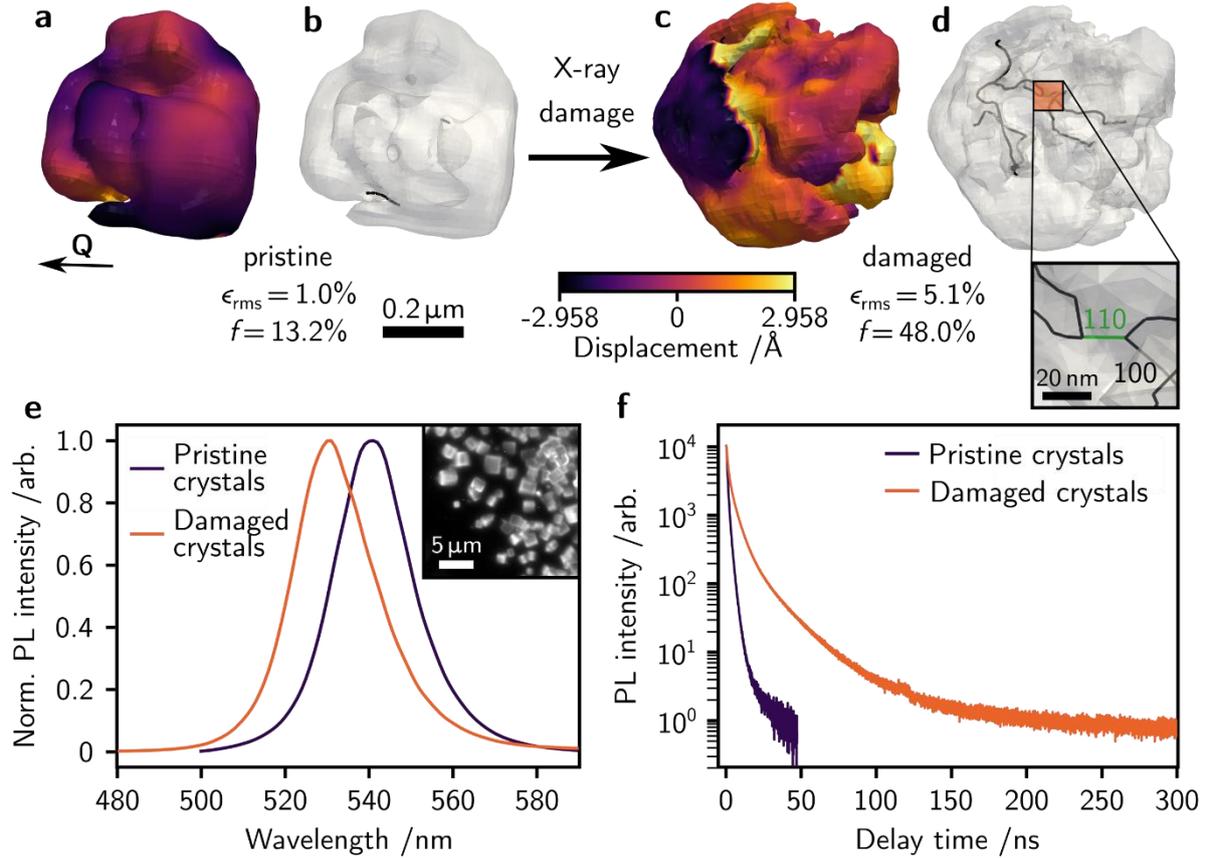

**Fig. 4: Impact of dislocation density on optoelectronic properties:** Electron density reconstructions of a crystal from its first BCDI scan (**a**, **b**), and from its second scan (**c**, **d**). Reconstructions are shown coloured according to the size of the atomic displacement along the direction of the scattering vector (**a**, **c**), and partially transparent in solid grey (**b**, **d**). Dislocations are shown as black lines. Exposure to X-rays damages the crystals, causing dislocation formation and increased strain (more dramatic changes in atomic displacement vector). The scale bar applies to all reconstructions. **e** PL spectra and **f** PL decays summed from pristine and damaged crystals upon excitation with a 509 nm laser using an excitation fluence of 63 nJcm$^{-2}$ for the pristine crystals measurement and 157 nJcm$^{-2}$ for the damaged crystals measurements.

**Comparisons to other semiconductors:**

Dislocation formation and migration are important mechanisms for strain relaxation in materials. For example, edge dislocations are known to form in epitaxial films of $CaCO_3$ under tensile stress beyond a critical layer thickness[47]. Considering that the vast majority of halide perovskite thin films are under tensile strain because of the high film annealing temperatures and the mismatch in thermal expansion properties between the perovskite and the substrate[48,49], we expect the light-induced formation and migration of these edge dislocations to be a strain relief mechanism active in working devices.

Further, there is rich dislocation formation and migration behaviour in semiconductors such as silicon and gallium pnictogenides, where dislocations also interact with localised (point) defects present in the crystals (though these studies do not consider the effect of visible light illumination)[50,51]. The face-centred cubic (FCC) structure is shared by Si, the anion sublattice of gallium pnictogenides, and the halide sublattice in many halide perovskites, with them each in turn sharing similar dislocation behaviour. Anderson *et al.*[30] note the stability of dislocations with a |**b**| of $\frac{1}{2}\langle 110 \rangle$ and $\langle 100 \rangle$ for FCC

systems in general. $\frac{1}{2}\langle 110\rangle$ dislocations are commonly observed in Si[52,53], GaN[54], GaP[55,56], and GaAs[57], and both $\frac{1}{2}\langle 110\rangle$ and $\langle 100\rangle$ dislocations have been found in FAPbI$_3$[12]. Here, we demonstrate that $\langle 100\rangle$ dislocations are common in MAPbBr$_3$ and that $\langle 110\rangle$ can also be found. Crucially, we also demonstrate the increased mobility of these dislocations under visible light illumination with nanometre resolution, giving unprecedented access to the buried structural changes occurring during device operation. Increased dislocation migration under infra-red irradiation has been reported in germanium crystals, but this was assessed by tracking the migration of dislocation rosettes on the surface of crystals that had been intentionally indented[58]. Our study uniquely tracks the migration of buried dislocations that result from common solution processed synthesis methods. Halide ions are the most mobile species in halide perovskites[36], therefore, one would expect them to be involved in dislocation migration. Given that the dislocations identified in this work are isostructural to those found in other FCC systems, we conclude that it is likely that the ions of the FCC halide sublattice move to realise dislocation migration. The mobility of dislocations in such high-performance semiconductors is surprising considering similar behaviour is seen in the highly disordered LiNi$_{0.5}$Mn$_{1.5}$O$_4$ battery electrode material during cycling[20], and calcite crystals during crystal growth and dissolution[19]. Ion movements with low activation barriers are fundamental to the electrochemistry and crystal growth processes at play in these studies, but are normally detrimental to the performance of traditional semiconductors in devices.

**CONCLUSION:**

By interrogating the atomic displacement fields present in crystals of MAPbBr$_3$ using BDCI, we have identified that $\langle 100\rangle$ edge dislocations are an important structural feature in halide perovskites used in optoelectronic devices, and that they are highly mobile under visible light illumination. Moreover, by intentionally studying a subset of crystals that are damaged by the X-ray beam, we discover that dislocation formation is a key feature of the halide perovskite degradation process, and that high dislocation densities correlate with a significant blue shift in the material's PL emission spectrum and a marked lengthening of its PL lifetime. Our results give a unique picture of the internal structure of halide perovskites and how it changes under device operation, elucidating the intimate links between nanoscale structure, dislocations, and device performance and stability.

Future work should focus on understanding the mechanism of dislocation migration and its relation to chemical and structural properties, including exploring why some halide perovskites show a greater propensity to degradation via dislocation formation than others. Computational calculations and atomic resolution electron microscopy have shown that in 2D WS$_2$, for example, highly mobile S atoms are instrumental in dislocation migration, and that this process has a remarkably low activation barrier.[59] Such an understanding of dislocation formation at the crystal growth stage will guide material and device fabrication procedures in order to avoid dislocation generation. We expect such *in situ* BCDI approaches with illumination could be used to study the evolution of intra-grain strain fields during device operation, dislocation-grain boundary interactions, and a variety of other photo-active semiconductor structural changes.

**METHODS:**

**Sample synthesis:**

Microcrystal film samples: Microcrystal film samples were synthesised by modifying the procedure reported by Saidaminov et al.[26,60]. The glass/patterned ITO substrate was cleaned by sonicating in detergent, de-ionized water, acetone, and isopropyl alcohol for 10 minutes each. The substrate was then dried with a nitrogen blow-dry and exposed to UV ozone treatment for 10 minutes. Methylammonium bromide solution was prepared by dissolving MABr and $PbBr_2$ in a 1:1 molar ratio in 5 ml DMF solvent. The 1 M concentration solution was filtered with 0.45 µm PTFE filter and mixed with 1,2-dichlorobenzene, DCB, in a 5:4 volumetric ratio (DMF: DCB) to obtain precipitated $MAPbBr_3$ crystals. The microcrystal films were obtained by pouring the solution onto prepared substrates inside a small-capped petri dish and submerging the substrates. The petri dish was covered with a lid and the solution stirred at 500 rpm for 30 minutes at 40 ˚C. The coated substrates were then removed and annealed at 110 ˚C for 5 min to leave a microcrystal film.

Isolated microcrystals: Samples of isolated microcrystals were synthesised using a variation on the antisolvent vapour assisted crystallisation technique. SiN membranes (Silson Ltd., product number: SiRN-10.0-200-3.0-200) were placed in a petri dish that was then set inside a large crystallisation basin containing chlorobenzene (Sigma Aldrich) antisolvent. 40 µL of precursor solution (0.1 M $PbBr_2$, 0.1 M MABr in 4:1 DMF:DMSO) were pipetted onto the SiN membrane, then the whole crystallisation basin was covered with aluminium foil to slow the evaporation of chlorobenzene by partially trapping the vapour in the basin before escape. The chlorobenzene vapour dissolves in the precursor solution at a slow rate, encouraging $MAPbBr_3$ crystals to crash out of solution.

**BCDI measurements:**

BCDI measurements were carried out at the I13-1 beamline of the Diamond Light Source (UK) using X-ray beam energies of 11.8 and 9.7 keV. Diffraction patterns were collected using the beamline's Excalibur photon-counting direct X-ray detector (Medipix3 chip) which was at a distance of 2.82 m from the sample. Measurements were taken in reflection geometry for microcrystals film samples, and were taken using a mixture of reflection and Laue (transmission) geometry for isolated crystal samples as these were on X-ray transparent SiN membranes. The crystals that became damaged under X-ray irradiation (Fig. 4) were measured using a beam energy of 9.7 keV. Data for reconstructions shown in all other figures were measured with an X-ray energy of 11.8 keV. At this energy, the proportion of crystals that suffered beam damage was smaller.

In a typical measurement, coherent diffraction patterns around the 100 Bragg peak were collected at 51 rocking curve angles separated by 0.005° (spanning a total angle range of 0.25°) with a collection dwell time of 10 s at each rocking curve angle.

To perform the measurements under illumination we constructed a home-built light-soaking rig on the beamline with a 405 nm continuous wave diode laser (CNI, model MLL-III-405) coupled to an optical fibre via a collimating lens. The laser power was tuned to achieve a *ca.* 1 sun intensity (in terms of photo-generated charge carriers) at the sample position. The power at this position was measured using a portable power meter (Thorlabs PM100D console unit; S120C Si photodiode power sensor).

**Electron density reconstruction:**

The measured coherent diffraction patterns were fast Fourier transformed back to real-space to get the crystal reconstructions. The amplitude was restored by taking the square-root of the intensity, while the phase was retrieved using iterative phasing methods. A linear combination of typical iterative phasing algorithms were used, including Error Reduction (ER), Hybrid Input-output (HIO)[22], and relaxed averaged alternating reflection (RAAR)[61] algorithms. The shrink-wrap[23] method was applied for updating the real-space constraints during the iterations and guided algorithms[62] were turned on for selecting the solution with minimum sharpness after each generation. Each diffraction pattern was reconstructed ten times with random initial guess to ensure reproducibility.

Reconstructions shown in this work are isovolumes whose surface is determined by setting a threshold value of electron density modulus and not displaying regions of space with modulus lower than this threshold value. For most reconstructions, a threshold of 0.1 was used (reconstructed electron density functions are normalised between 0 and 1).

**Determination of Burgers vector magnitudes:**

Electron density reconstructions were produced in .vtk file format and viewed and analysed using the open-source Paraview[63] data visualisation software. Slices through the reconstruction were taken perpendicular (by eye) to the dislocation line and then displacement values were extracted along circular paths centred on the dislocation core at three different radii, $r$. Note that $r$ should be small enough so the strain field from the dislocation dominates any contribution from other strain fields present in the crystal, but large enough not to be on the dislocation core. The resulting displacement *vs.* arc angle data was fit according to the function given in the main text using the SciPy Python package. The three calculated Burgers vector magnitudes were then averaged.

**Hyperspectral PL microscopy:**

A Photon Etc. IMA Vis microscope was used for the wide field hyperspectral microscopy studies with 100X Olympus and 50X Nikon aberration corrected objective lenses in place. Sample excitation was achieved using a 405 nm continuous wave laser filtered with a dichroic mirror. Spectral resolution is achieved since the emitted light from the sample is spectrally split through a volume Bragg grating before being collected by charged-coupled device (CCD) camera. The camera used is a 2048 x 2048 resolution Hamamatsu Orca Flash V3.0 with a wavelength range of 400-1000 nm and is maintained at -10 °C during measurements. The position on the sample from which light is emitted is calculated by scanning the angle of the grating relative to the emitted light.

In a typical measurement, PL is collected from an area of 89 μm × 89 μm using a (wide field) incident laser intensity of 30 μW/cm$^2$ and a dwell time of 10 s at each collection wavelength.

**Confocal PL microscopy:**

Fluorescence lifetime imaging maps were acquired with a MicroTime200 laser scanning confocal microscope set-up from Picoquant. Fluorescence was excited with a 509 nm diode laser (LDH-D-C-510, Picoquant; 2.96 MHz) using a 100X 0.9NA objective (Olympus MPlanFL N). The PL was collected in reflection geometry, sent through a pinhole of 100 μm diameter, and detected with an Excelitas SPAD detector. The system has an approximate time resolution of 400 ps. Unwanted scattering from the excitation laser is suppressed by a combination of a long-pass dichroic mirror and a 519 nm long-pass filter.

In a typical measurement, an area of 10 x 10 μm was divided into 256 x 256 pixels with a dwell time of 10 μs. The eventual image is calculated as the sum of 200 complete scans. For the measurements shown in Fig. 4f, a laser power of 4.7 nW was used with a repetition rate of 20 MHz for the pristine crystals, and 8 MHz for the damaged crystals. The spot diameter is estimated using the airy disc diameter of $1.22 \times \frac{\lambda}{NA}$.

**AUTHOR CONTRIBUTIONS:**

S.D.S., I.K.R., T.A.S.D, and K.W.P.O. conceived of the project. K.W.P.O, J.D., D.J.B, A.N.I., S.K., K.F., M.D., S.J.Z, A.E.D., T.A.S., and P.L. carried out the BCDI measurements. J.D. generated the crystal reconstructions, which were analysed by K.W.P.O. K.W.P.O. carried out the hyperspectral and confocal photoluminescence microscopy measurements. M.N.L. prepared the microcrystal film samples and K.W.P.O. prepared the isolated microcrystal samples. A.E.D. synthesised the large flat microcrystal samples.

S.D.S. supervised K.W.P.O., A.N.I., K.F., T.A.S., and T.A.S.D. I.K.R. supervised J.D. O.B. supervised M.N.L. S.D.S. and S.H. supervised A.E.D.

K.W.P.O. produced first drafts and all authors contributed to editing the manuscript.

**ACKNOWLEDGEMENTS:**

K.W.P.O. acknowledges an EPSRC studentship. J.D. thanks the China Scholarship Council (CSC) for financial support. M.N.L and O.M.B. acknowledge financial support from KAUST. A.N.I. acknowledges scholarships from the British Spanish Society, Sir Richard Stapley Educational Trust, and the Rank Prize Fund. S.K. acknowledges funding from the Leverhulme Trust (ECF-2022-593), the Isaac Newton Trust (22.08(i)), and the German Academic Foreign Service (91793256). K.F. acknowledges a Winton Sustainability Fund Studentship, a George and Lilian Schiff Studentship and an Engineering and Physical Sciences Research Council (EPSRC) studentship. M.D. acknowledges Leverhulme Research Grant RPG-2021-191. J. Z. acknowledges support from the Polish National Agency for Academic Exchange within the Bekker programme (grant no. PPN/BEK/2020/1/00264/U/00001). A.E.D. Acknowledges funding from EPSRC Cambridge NanoDTC, EP/L015978/1. T.A.S. Acknowledges funding from EPSRC Cambridge NanoDTC, EP/S022953/1. T.A.S.D. acknowledges the support of an Ernest Oppenheimer Early Career Fellowship and a Schmidt Science Fellowship. S.H. acknowledges funding from EPSRC (EP/T001038/1). I.K.R. acknowledges support from the U.S. Department of Energy, Office of Science, Office of Basic Energy Sciences, under Contract No. DE-SC0012704 and EPSRC. S.D.S. acknowledges the Royal Society and Tata Group (grant no. UF150033). The work has received funding from the European Research Council under the European Union's Horizon 2020 research and innovation programme (HYPERION, grant agreement no. 756962). The authors acknowledge the EPSRC (EP/R023980/1, EP/S030638/1) for funding. The authors acknowledge Diamond Light Source for time on Beamline I13-1 under proposal numbers MG25097-1, MG28495-1, and MG30308-1.

Work at Brookhaven National Laboratory was supported by the U.S. Department of Energy, Office of Science, Office of Basic Energy Sciences, under Contract No. DE-SC0012704. Work performed at UCL was supported by EPSRC. For the purpose of open access, the author has applied a Creative Commons Attribution (CC BY) licence to any Author Accepted Manuscript version arising.

**COMPETING INTERESTS:**

S.D.S. is a cofounder of Swift Solar.

# Imaging Light-Induced Migration of Dislocations in Halide Perovskites with 3D Nanoscale Strain Mapping

## SUPPLEMENTARY INFORMAITON

Kieran W. P. Orr[1,2], Jiecheng Diao[3,4], Muhammad Naufal Lintangpradipto[5], Darren J. Batey[6], Affan N. Iqbal[1,2], Simon Kahmann[1,2], Kyle Frohna[1], Milos Dubajic[2], Szymon J. Zelewski[1,2], Alice E. Dearle[1,2,7], Thomas A. Selby[2], Peng Li[6], Tiarnan A. S. Doherty[1,2,8], Stephan Hofmann[7], Osman M. Bakr[5], Ian K. Robinson[3,9], Samuel D. Stranks*[1,2] [1]*Department of Physics, Cavendish Laboratory, University of Cambridge, JJ Thomson Avenue, Cambridge, CB3 0HE, UK*
[2]*Department of Chemical Engineering and Biotechnology, University of Cambridge, Philippa Fawcett Drive, Cambridge, CB3 0AS, UK*
[3]*London Centre for Nanotechnology, University College London, London WC1E 6BT, UK*
[4]*Center for Transformative Science, ShanghaiTech University, Shanghai 201210, China*
[5]*Physical Science and Engineering (PSE) Division, King Abdullah University of Science and Technology (KAUST) Thuwal 23955-6900, Kingdom of Saudi Arabia*
[6]*Diamond Light Source, Harwell Science and Innovation Campus, Fermi Ave, Didcot OX11 0DE, UK*
[7]*Department of Engineering, University of Cambridge, UK*
[8]*Department of Materials Science & Metallurgy, University of Cambridge, 27 Charles Babbage Road, Cambridge, CB3 0FS, United Kingdom*
[9]*Condensed Matter Physics and Materials Science Department, Brookhaven National Lab, Upton, New York 11793, USA*

*sds65@cam.ac.uk

This work was carried out while J.D. was affiliated with [3]. J.D. is now affiliated with [4].

**SUPPLEMENTARY TABLE:**

| Where does the reconstruction appear? | Fraction of crystal volume with local strain greater than 1%, $f$ /% | Root mean squared local strain, $\epsilon_{\text{rms}}$ /% |
| --- | --- | --- |
| Main text Fig. 1b | 13.2 | 0.7 |
| Main text Fig. 2a & Supplementary Fig. 2a | 18.7 | 2.9 |
| Supplementary Fig. 2b | 23.6 | 3.1 |
| Supplementary Fig. 2c | 27.1 | 4.3 |
| Main text Fig. 3a & Supplementary Fig. 2d | 27.9 | 3.9 |
| Main text Fig. 3b | 14.9 | 3.6 |
| Main text Fig. 3c | 25.0 | 3.6 |
| Main text Fig. 3d | 23.0 | 3.8 |
| Main text Fig. 3e | 21.2 | 3.5 |
| Main text Fig. 4a & 4b | 13.2 | 1.0 |
| Main text Fig. 4c & 4d (beam damaged) | 48.0 | 5.1 |
| Supplementary Fig. 24a & 24b | 6.6 | 0.6 |
| Supplementary Fig. 24c & 24d (beam damaged) | 14.2 | 1.9 |

**Supplementary Table 1: Local strain in MAPbBr$_3$ microcrystals:** By taking the spatial derivative of the atomic displacement field between voxels of the reconstructions with respect to the X-ray scattering vector direction we calculate the local nanoscale tensile (positive)/compressive (negative) strain in a crystal. These values can be plotted as a histogram, such as the one shown in Fig. 1e of the main text. We then calculate the fraction of points where the local tensile/compressive strain is greater than 1% (shown by the orange regions of the histogram in Fig. 1e of the main text). We also tabulate the root mean square local strain within the crystals.

Histograms of the underlying local strain distributions are shown at the end of the Supplementary Figures (apart from that for the crystal in Fig. 1b of the main text which is shown in Fig. 1e of the main text).

**SUPPLEMENTARY NOTE 1:**

**Missing reconstruction volume due to crystal structure twisting:**

Where reconstructions are not well faceted, the missing volume is due to parts of the crystal that do not satisfy the Bragg condition for the position of the detector at the beamline. For example, such regions of a crystal could be twisted into another orientation. We find evidence for such twisting in related $MAPbBr_3$ crystals which are large and flat, and are synthesised in an analogous fashion to the crystals considered in the main text. Supplementary Fig. 1a and 1b show two maps of diffracted X-ray intensity from one of these large, flat $MAPbBr_3$ crystals. The difference in incidence angle between the sample and X-ray beam for these two maps is 0.05°. Regions (outlined in orange) of high diffracted intensity in Supplementary Fig. 1b align with regions of low diffracted intensity in Supplementary Fig. 1a indicating that these regions of the crystal must be misoriented from each other by 0.05°. This large, flat crystal of $MAPbBr_3$ is contiguous and the shape of the crystal does not give obvious reasons for the lattice misorientation within its structure (Supplementary Fig. 1c).

Therefore, any lack of faceting in the reconstructions shown in the main text is likely due to misalignment of the crystalline lattice towards the edges of a well-facetted crystal (Fig. 1c & d of the main text confirm the well-facetted nature of our sample crystals), rather than any errors in reconstruction. Additionally, reconstructions were performed ten times with random initial starting guesses of diffraction pattern phases and real-space support volume and were checked for consistency.

Another possible reason for any missing volumes in the reconstruction could be due to the presence of point defects and vacancies in the samples which lower the overall scattering power of the crystals, resulting in smaller reconstruction volumes.

The large flat single crystals of $MAPbBr_3$ such as the one considered here are made according to the following synthesis procedure (closely related to the procedures used to make the samples considered in the main text). A $MAPbBr_3$ precursor solution of methylammonium bromide (Great Cell Solar, >99.99%) and lead (II) bromide (TCI, >98.0%) in N,N-dimethylformamide (Sigma-Aldrich, 99.8%) was prepared with concentration 1 M. A droplet of very limited volume (<< 1μL) of MAPbBr3 precursor solution, was deposited onto a silicon wafer coated in 10 μm of $SiO_2$. The precursor droplet is rapidly confined with an exfoliated flake of biotite mica. After *ca.* 72 hours; space-confined perovskite crystallisation occurs by evaporation and the biotite mica flake can be easily removed.

**SUPPLEMENTARY NOTE 2:**

**Control scans for crystal in Fig. 3 of the main text in dark conditions:**

In order to confirm that dislocations migrate more easily under visible light illumination, we first needed to collect BCDI data in dark conditions (i.e. with only the beamline hutch ambient lights on). For the crystal considered in Fig. 3 of the main text, we first measured four BCDI scans without laser illumination. These reconstructions are shown in Supplementary Fig. 2. The overall shape of the reconstruction is largely reproducible and the dislocation remains along the long-axis of the reconstruction. Under illumination the dislocations become highly mobile, as shown in Fig. 3 of the main text.

Further, we can be confident the increased dislocation migration is caused by the visible light illumination, as opposed to X-ray beam damage, because the crystal is radiation stable as confirmed by the plots of Bragg peak intensity as a function of rocking curve angle for successive scans shown in Supplementary Fig. 3. We can also be confident that the crystals are not being damaged by the X-ray beam because the region of sample that was exposed to the X-ray beam for the scans leading to the reconstructions in Fig. 3 of the main text shows no difference in PL spectrum compared to a nearby region of sample that was not exposed to the beam as is shown in Supplementary Fig. 4.

A beam energy of 11.8 keV was used for the Bragg CDI scans for Fig. 3. This X-ray energy was far less damaging than 9.7 keV which was used for the measurements where we saw definite beam damage (Fig. 4 of the main text and Supplementary Fig. 24).

**SUPPLEMENTARY FIGURES:**

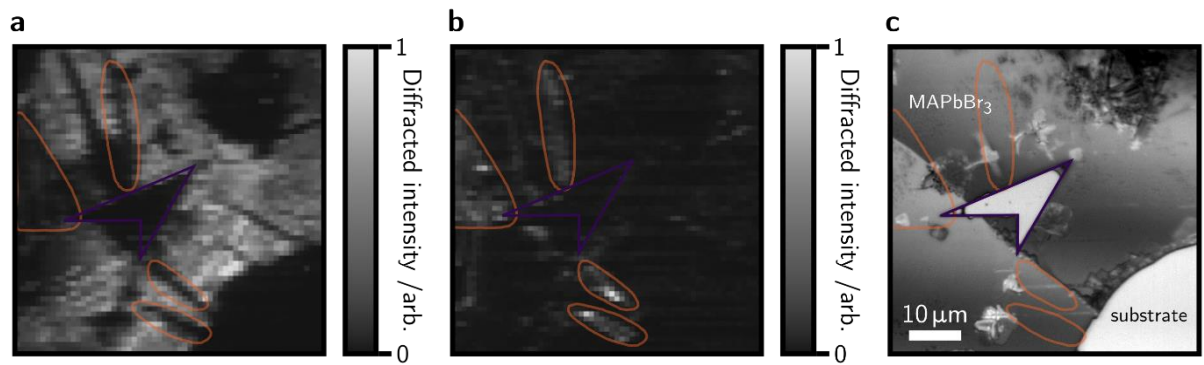

**Supplementary Fig. 1: Misorientation within a MAPbBr$_3$ single crystal: a** and **b** show maps of diffracted intensity as the X-ray beam is raster scanned across the sample. The difference in rocking angle at which the sample was held for scan **a** and **b** is 0.05°. **c** shows a broadband optical microscopy image of this same region of crystal (as can be identified by the arrow-head shape highlighted in purple).

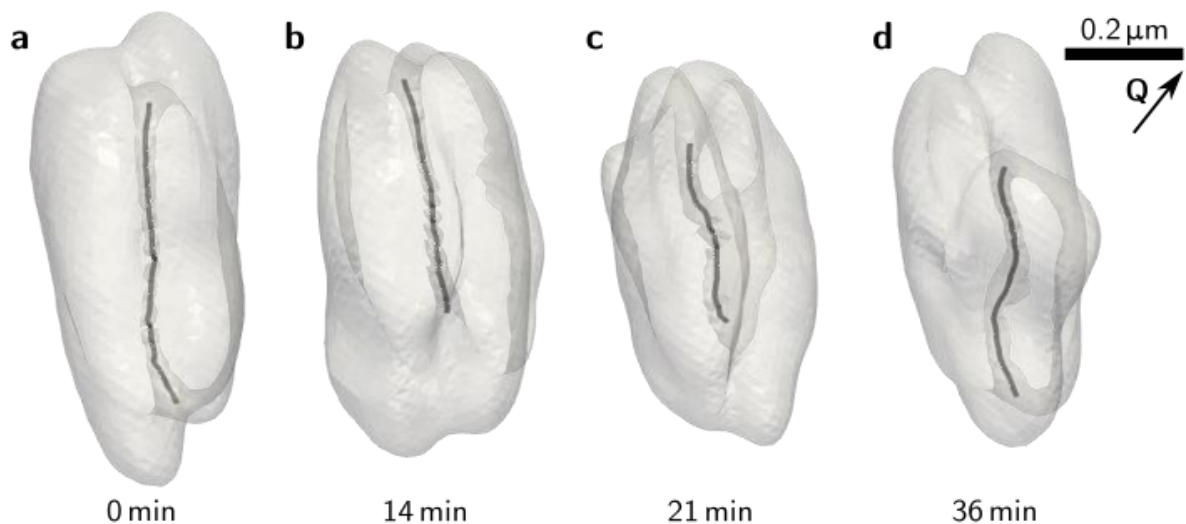

**Supplementary Fig. 2: Control reconstructions for crystal considered in Fig. 3 of the main text:** Reconstructions of successive Bragg CDI scans for the crystal considered in Fig. 3 of the main text. Four scans were carried out on the crystal before illuminating with the 405nm laser to check that the crystal was stable to the X-ray radiation. **a**, **b**, **c**, and **d** show the reconstructions from these scans with the dislocation highlighted in black. The time given under the reconstruction is the time of the scans relative to the first (**a**). The scale bar and scattering vector apply to all reconstructions. The reconstruction shown in panel **d** is the same as that shown in Fig. 3a of the main text. Burgers vector characterisation for the dislocations in these crystals confirming that they are all ⟨100⟩ edge dislocations are shown in Supplementary Fig. 5-8.

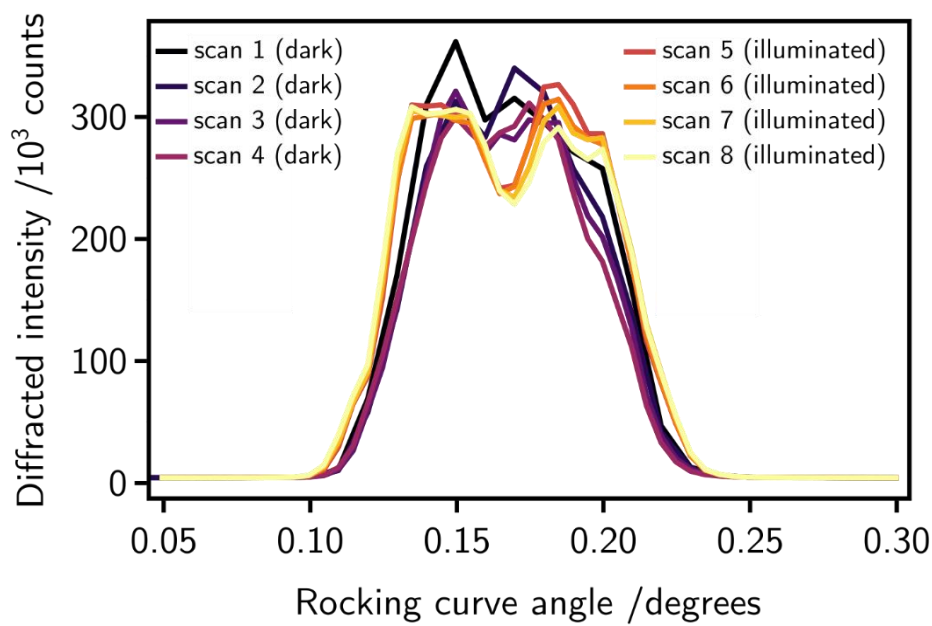

**Supplementary Fig. 3: Radiation stability of crystal considered in Supplementary Fig. 2 and Fig. 3 of the main text:** Bragg peak intensity as a function of rocking curve angle for successive scans of the MAPbBr$_3$ microcrystal considered in Supplementary Fig. 2 and Fig. 3 of the main text. The Bragg peak intensity and general shape is constant, indicating negligible beam damage is taking place during each successive scan for this crystal.

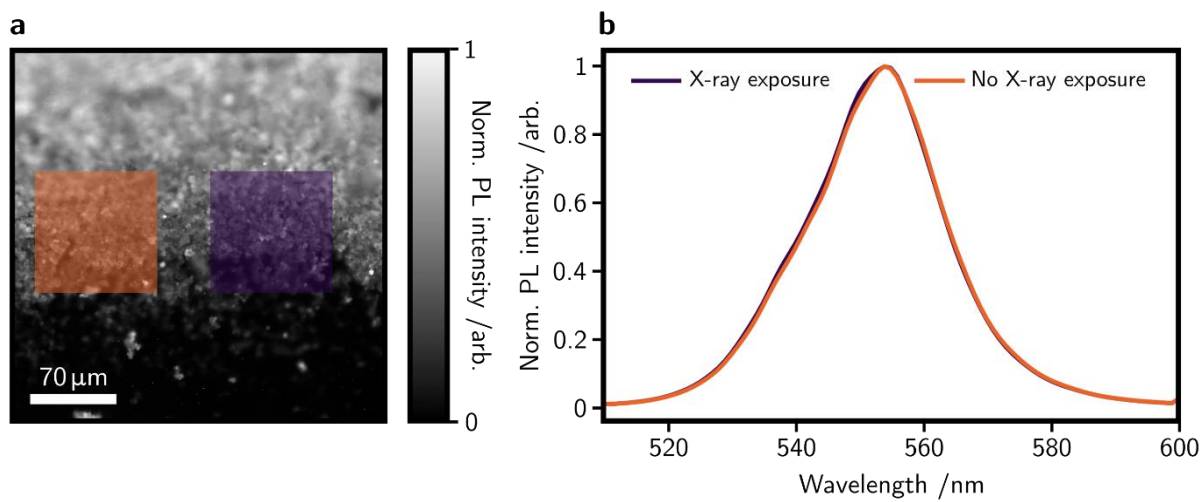

**Supplementary Fig. 4: PL stability of crystals under 11.8keV X-rays: a** Map of normalised PL emission intensity of a microcrystal film. The black region at the bottom of the image is where the film ends and the top of the image is out of focus because the edge of the film thins gradually rather than abruptly. The centre of the image is in focus at a mid-point of the microcrystal film thickness. The region highlighted in purple was exposed to X-rays and the region highlighted in orange was not. **b** PL emission spectra from the purple and orange regions in **a**. No significant differences in PL emission spectra are observed, further confirming lack of beam damage.

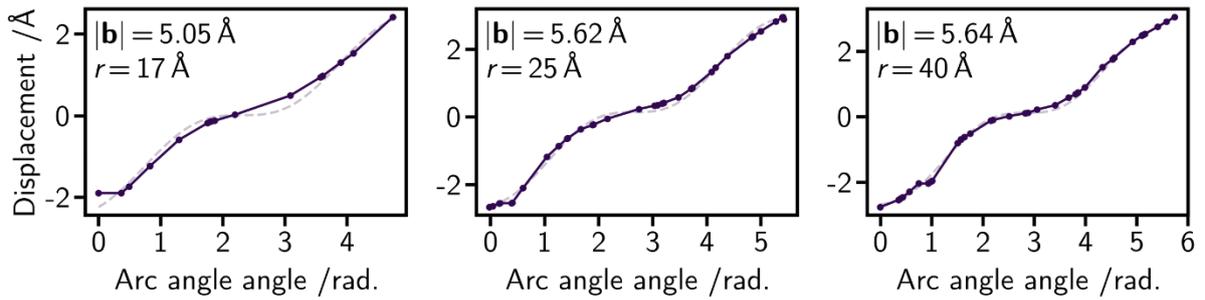

**Supplementary Fig. 5: Burgers vector calculation for Supplementary Fig. 2a (dark):** Atomic displacements as a function of arc angle as we circle the dislocation core at three different radii. Dashed lines: fit to data of function for atomic displacement, $u$, given in the main text. Average Burgers vector magnitude is 5.44Å.

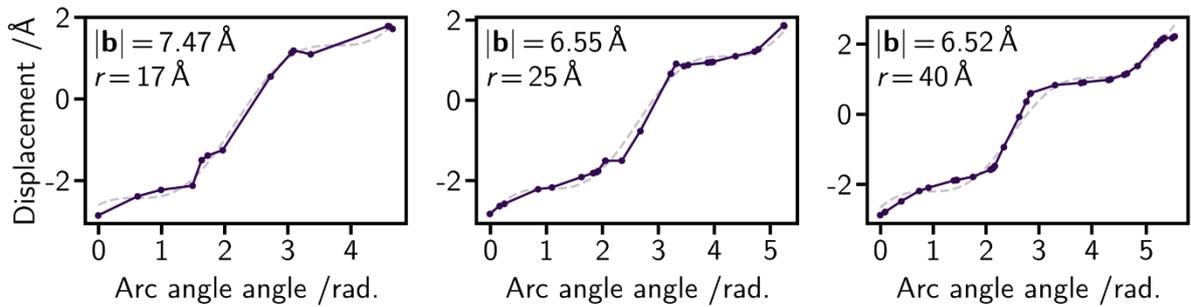

**Supplementary Fig. 6: Burgers vector calculation for Supplementary Fig. 2b (dark):** Atomic displacements as a function of arc angle as we circle the dislocation core at three different radii. Dashed lines: fit to data of function for atomic displacement, $u$, given in the main text. Average Burgers vector magnitude is 6.84Å.

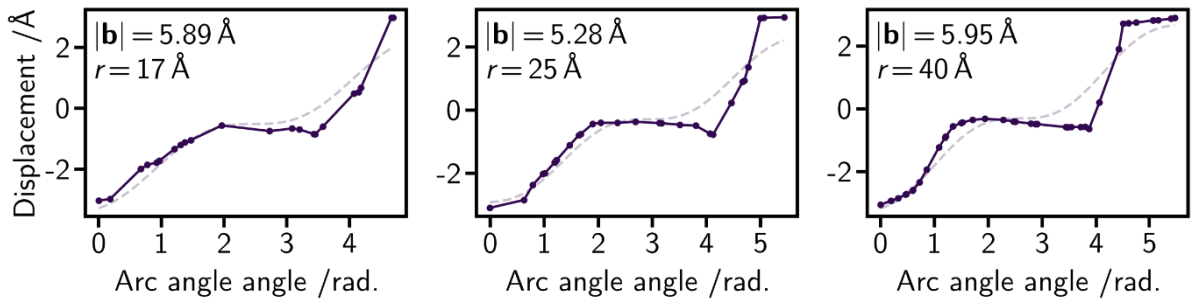

**Supplementary Fig. 7: Burgers vector calculation for Supplementary Fig. 2c (dark):** Atomic displacements as a function of arc angle as we circle the dislocation core at three different radii. Dashed lines: fit to data of function for atomic displacement, $u$, given in the main text. Average Burgers vector magnitude is 5.71Å.

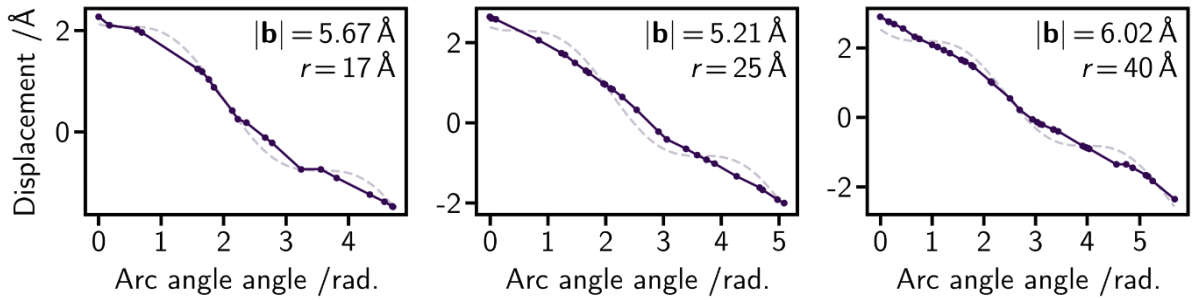

**Supplementary Fig. 8: Burgers vector calculation for Fig. 3a of the main text and Supplementary Fig. 2d (dark):** Atomic displacements as a function of arc angle as we circle the dislocation core at three different radii. Dashed lines: fit to data of function for atomic displacement, $u$, given in the main text. Average Burgers vector magnitude is 5.97Å.

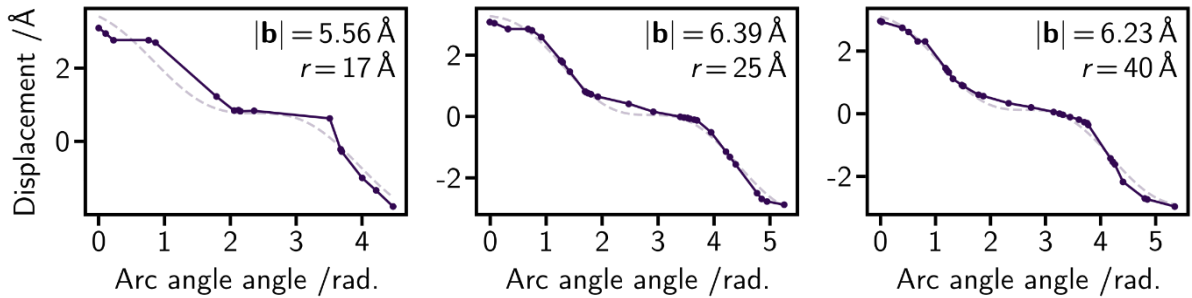

**Supplementary Fig. 9: Burgers vector calculation for Fig. 3b of the main text (illuminated):** Atomic displacements as a function of arc angle as we circle the dislocation core at three different radii. Dashed lines: fit to data of function for atomic displacement, $u$, given in the main text. Average Burgers vector magnitude is 6.06Å. Illumination time = 0 min (after 105 min of scan parameter re-optimisation under illumination).

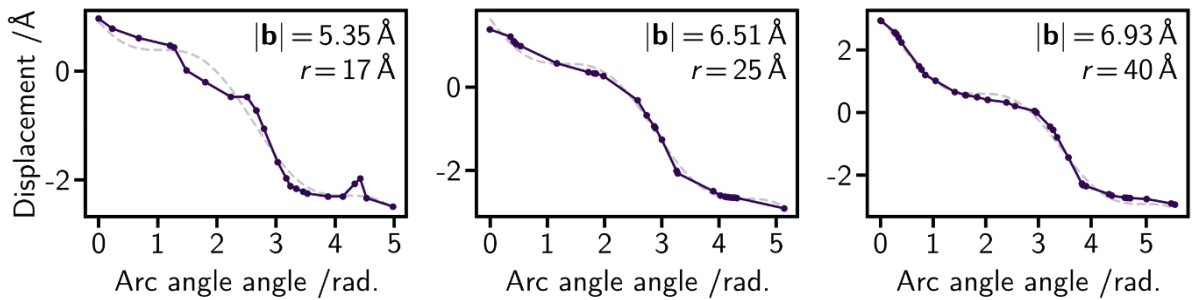

**Supplementary Fig. 10: Burgers vector calculation for Fig. 3c of the main text:** Atomic displacements as a function of arc angle as we circle the dislocation core at three different radii. Dashed lines: fit to data of function for atomic displacement, $u$, given in the main text. Average Burgers vector magnitude is 6.26Å. Illumination time = 14 min (after 105 min of scan parameter re-optimisation under illumination).

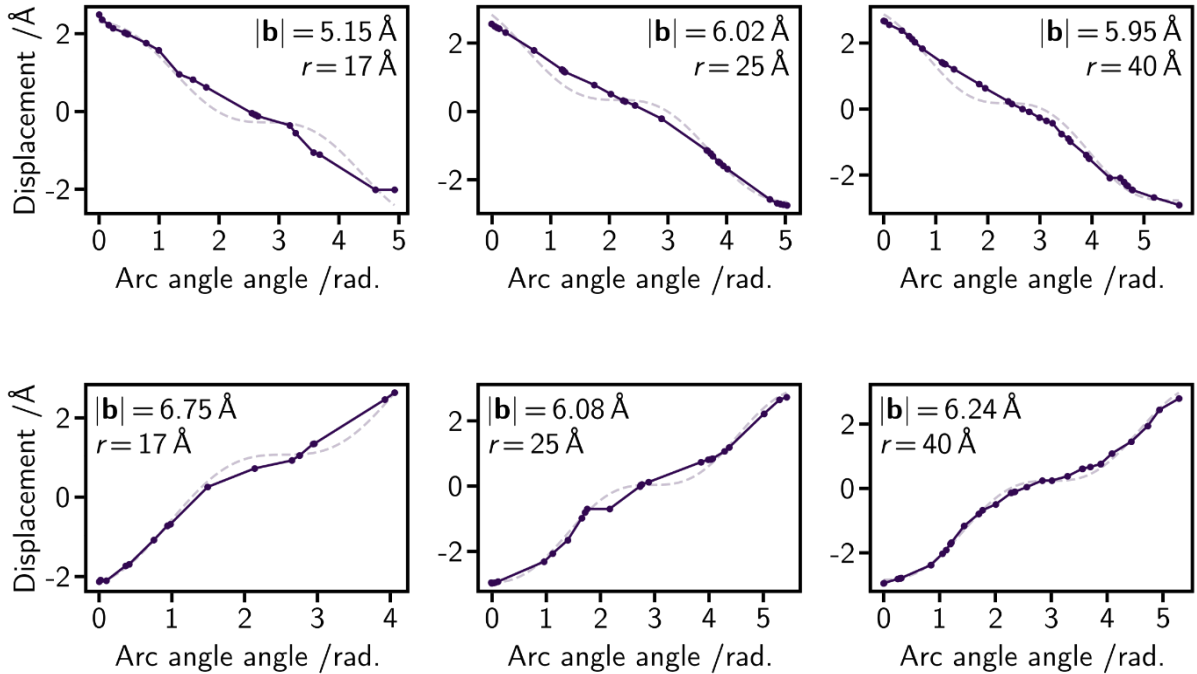

**Supplementary Fig. 11: Burgers vector calculation for Fig. 3d of the main text:** Atomic displacements as a function of arc angle as we circle the dislocation core at three different radii. Dashed lines: fit to data of function for atomic displacement, $u$, given in the main text. Average Burgers vector magnitudes are 5.70Å (long dislocation) and 6.36Å (short dislocation). Illumination time = 27 min (after 105 min of scan parameter re-optimisation under illumination).

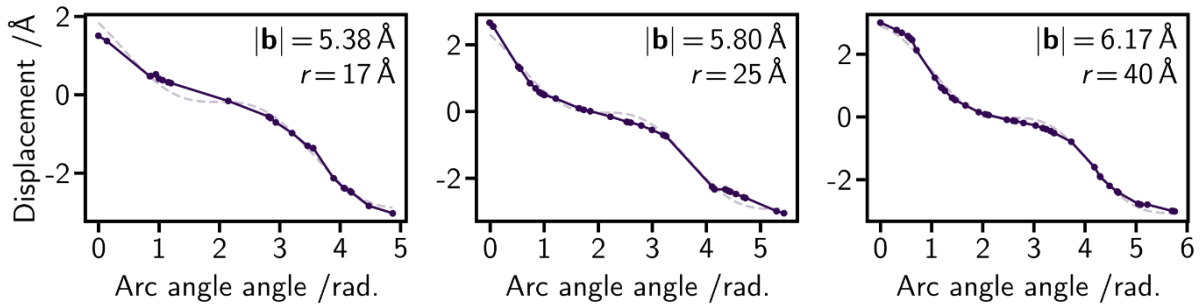

**Fig. 12: Burgers vector calculation for Fig. 3e of the main text:** Atomic displacements as a function of arc angle as we circle the dislocation core at three different radii. Dashed lines: fit to data of function for atomic displacement, $u$, given in the main text. Average Burgers vector magnitude is 5.78Å. Illumination time = 37 min (after 105 min of scan parameter re-optimisation under illumination).

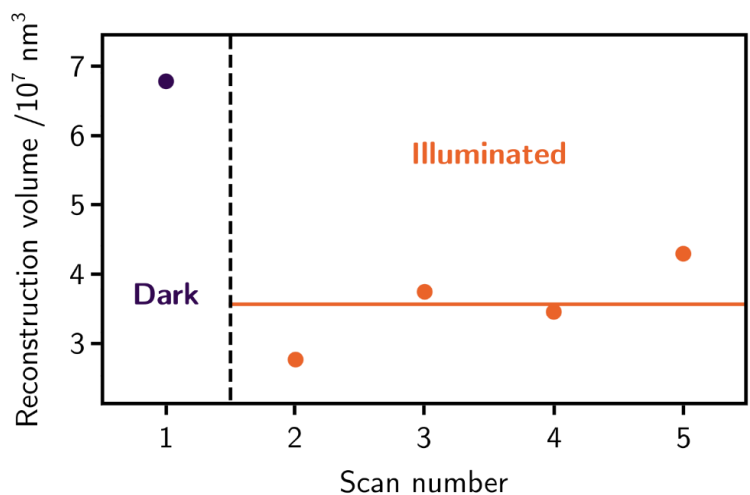

**Fig. 13: Volumes of reconstructions shown in Fig. 3 of the main text:** Values given are calculated using an isovolume threshold value of 0.16 (maximum electron density normalised to 1). The horizontal line shows the average volume for the reconstructions in when measuring under illumination.

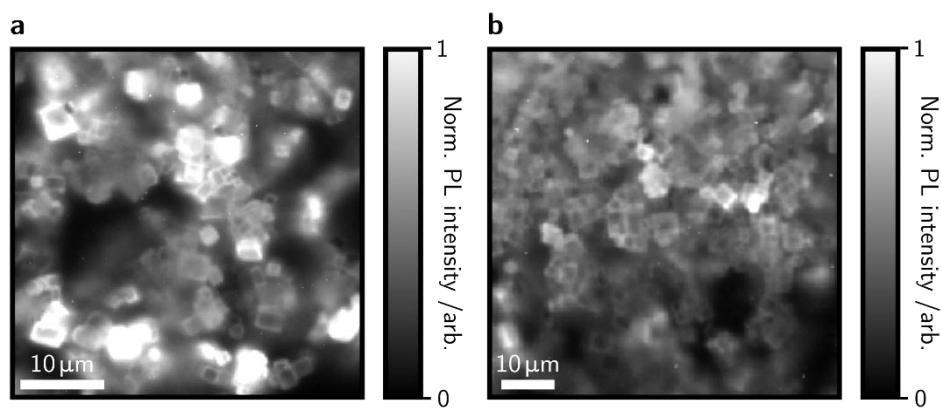

**Supplementary Fig. 14: PL microscopy:** Total emitted PL intensity images (405nm excitation wavelength) of **a** pristine MAPbBr$_3$ microcrystals, and **b** microcrystals in a region of a microcrystal film sample that was exposed to X-rays. Measurements were carried out using a hyperspectral microscope according to the method given in the main text. It can be seen that the crystals are well facetted before and after X-ray exposure.

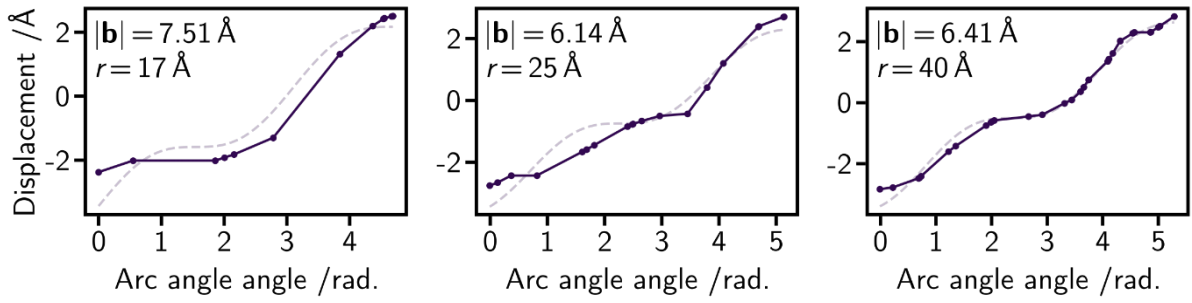

**Supplementary Fig. 15: Burgers vector calculation for crystal in Fig. 4a and b of the main text, long dislocation:** Atomic displacements as a function of arc angle as we circle the dislocation core at three different radii. Dashed lines: fit to data of function for atomic displacement, $u$, given in the main text. Average Burgers vector magnitude is 6.69Å.

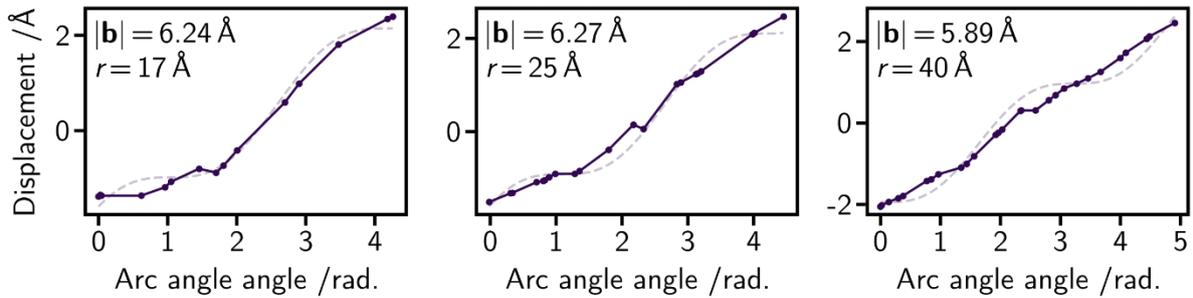

**Supplementary Fig. 16: Burgers vector calculation for crystal in Fig. 4a and b of the main text, short dislocation:** Atomic displacements as a function of arc angle as we circle the dislocation core at three different radii. Dashed lines: fit to data of function for atomic displacement, $u$, given in the main text. Average Burgers vector magnitude is 6.13Å.

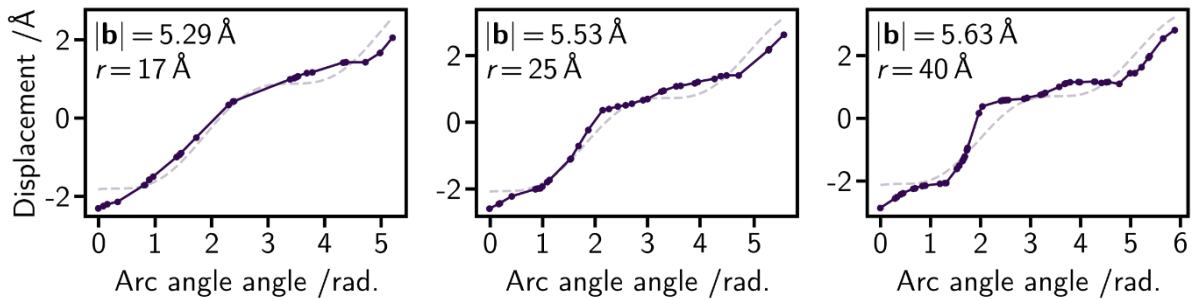

**Supplementary Fig. 17: Burgers vector calculation for crystal in Fig. 4c and d of the main text, dislocation 1:** Atomic displacements as a function of arc angle as we circle the dislocation core at three different radii. Dashed lines: fit to data of function for atomic displacement, $u$, given in the main text. Average Burgers vector magnitude is 5.48Å.

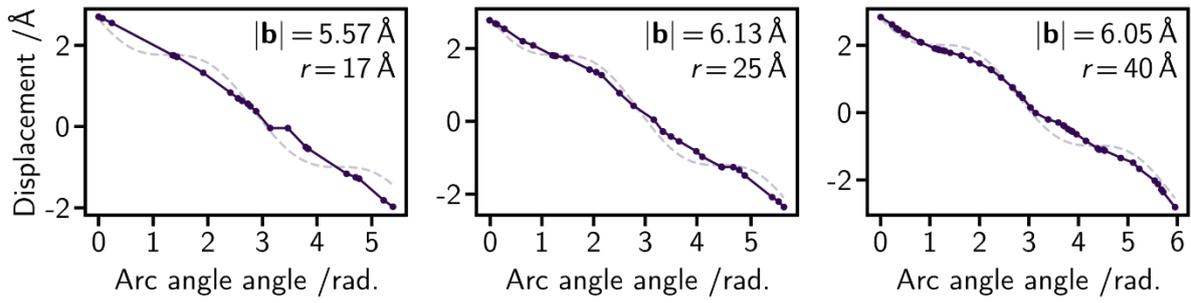

**Supplementary Fig. 18: Burgers vector calculation for crystal in Fig. 4c and d of the main text, dislocation 2:** Atomic displacements as a function of arc angle as we circle the dislocation core at three different radii. Dashed lines: fit to data of function for atomic displacement, $u$, given in the main text. Average Burgers vector magnitude is 5.92Å.

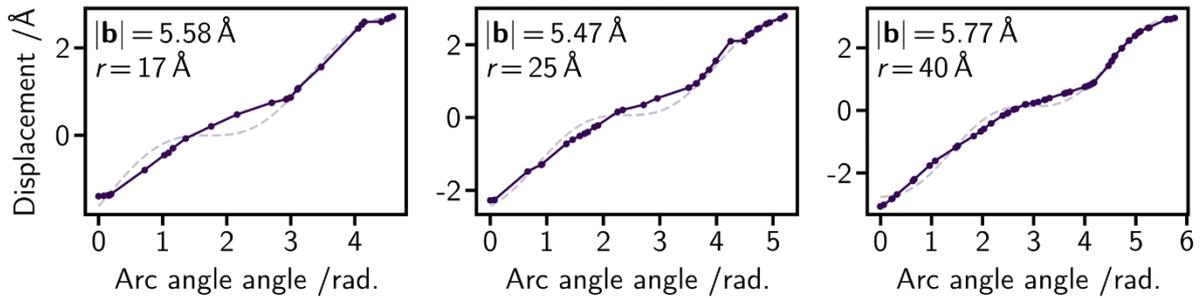

**Supplementary Fig. 19: Burgers vector calculation for crystal in Fig. 4c and d of the main text, dislocation 3:** Atomic displacements as a function of arc angle as we circle the dislocation core at three different radii. Dashed lines: fit to data of function for atomic displacement, $u$, given in the main text. Average Burgers vector magnitude is 5.61Å.

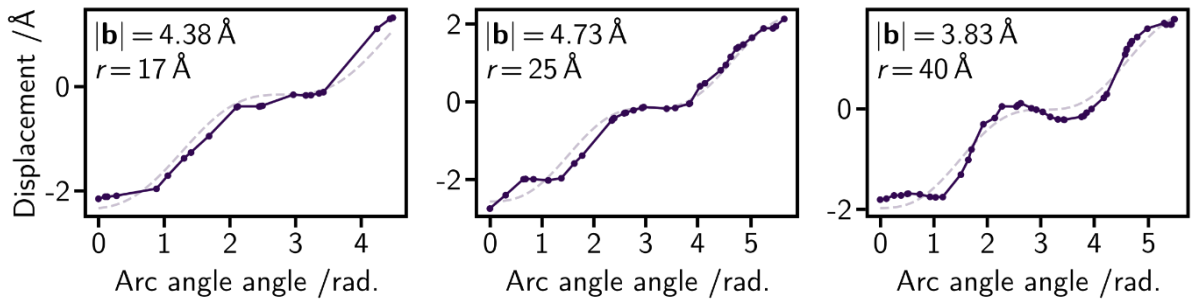

**Supplementary Fig. 20: Burgers vector calculation for crystal in Fig. 4c and d of the main text, dislocation 4:** Atomic displacements as a function of arc angle as we circle the dislocation core at three different radii. Dashed lines: fit to data of function for atomic displacement, $u$, given in the main text. Average Burgers vector magnitude is 4.31Å. This is the dislocation with a Burgers vector $\simeq d_{110}$.

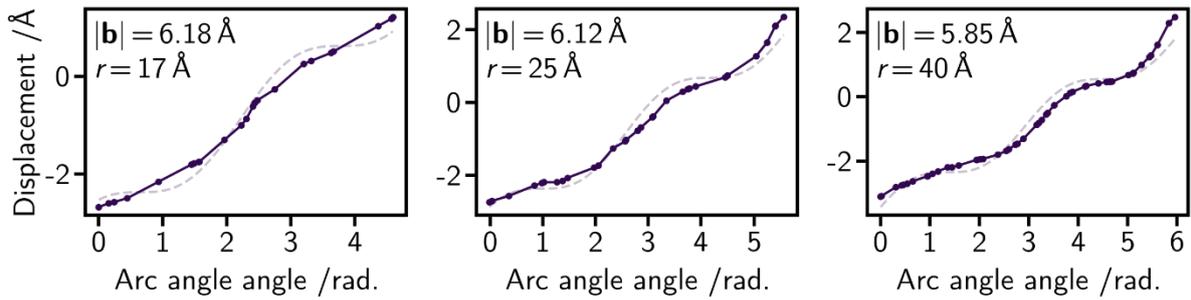

**Supplementary Fig. 21: Burgers vector calculation for crystal in Fig. 4c and d of the main text, dislocation 5:** Atomic displacements as a function of arc angle as we circle the dislocation core at three different radii. Dashed lines: fit to data of function for atomic displacement, $u$, given in the main text. Average Burgers vector magnitude is 6.05Å.

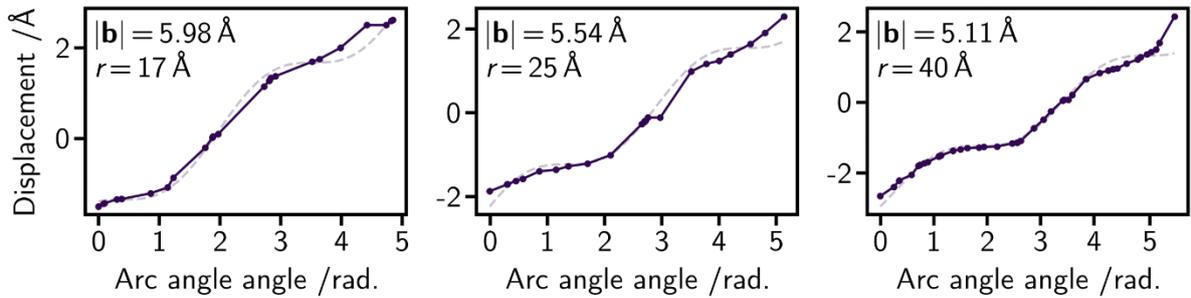

**Supplementary Fig. 22: Burgers vector calculation for crystal in Fig. 4c and d of the main text, dislocation 6:** Atomic displacements as a function of arc angle as we circle the dislocation core at three different radii. Dashed lines: fit to data of function for atomic displacement, $u$, given in the main text. Average Burgers vector magnitude is 5.54Å.

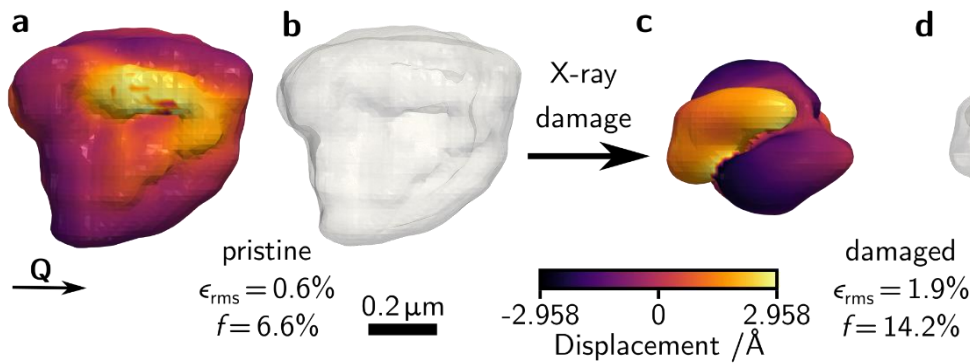

**Supplementary Fig. 23: Additional example of X-ray induced damage and dislocation formation:** Electron density reconstructions a crystal from its first BCDI scan (**a**, **b**), and from a second scan (**c**, **d**). Reconstructions are shown coloured according to the size of the atomic displacement along the direction of the scattering vector (**a**, **c**), and partially transparent in solid grey (**b**, **d**). Dislocations are shown as black lines. Exposure to X-rays damages the crystals, causing dislocation formation and increased strain (more dramatic changes in atomic displacement vector) as well as reduced crystallite volume. The scale bar and scattering vector apply to all reconstructions. Displacement *vs.* arc angle plots for the dislocation formed and the associated Burgers vectors are shown in Supplementary Fig. 25.

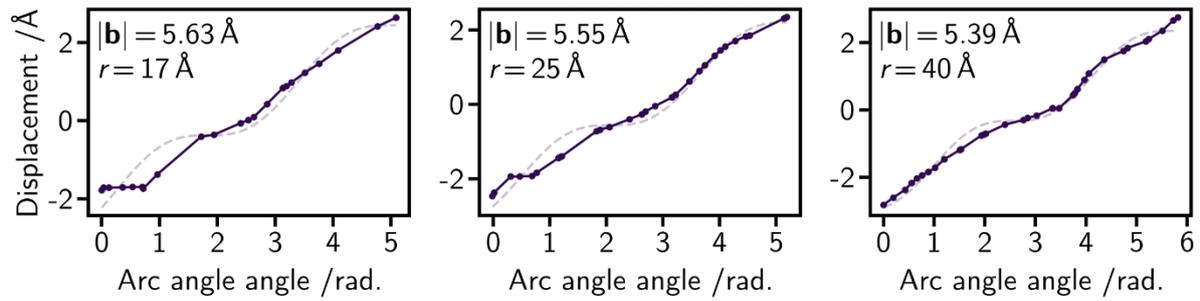

**Supplementary Fig. 24: Burgers vector calculation for crystal in Supplementary Fig. 23:** Atomic displacements as a function of arc angle as we circle the dislocation core at three different radii. Dashed lines: fit to data of function for atomic displacement, $u$, given in the main text. Average Burgers vector magnitude is 5.52Å.

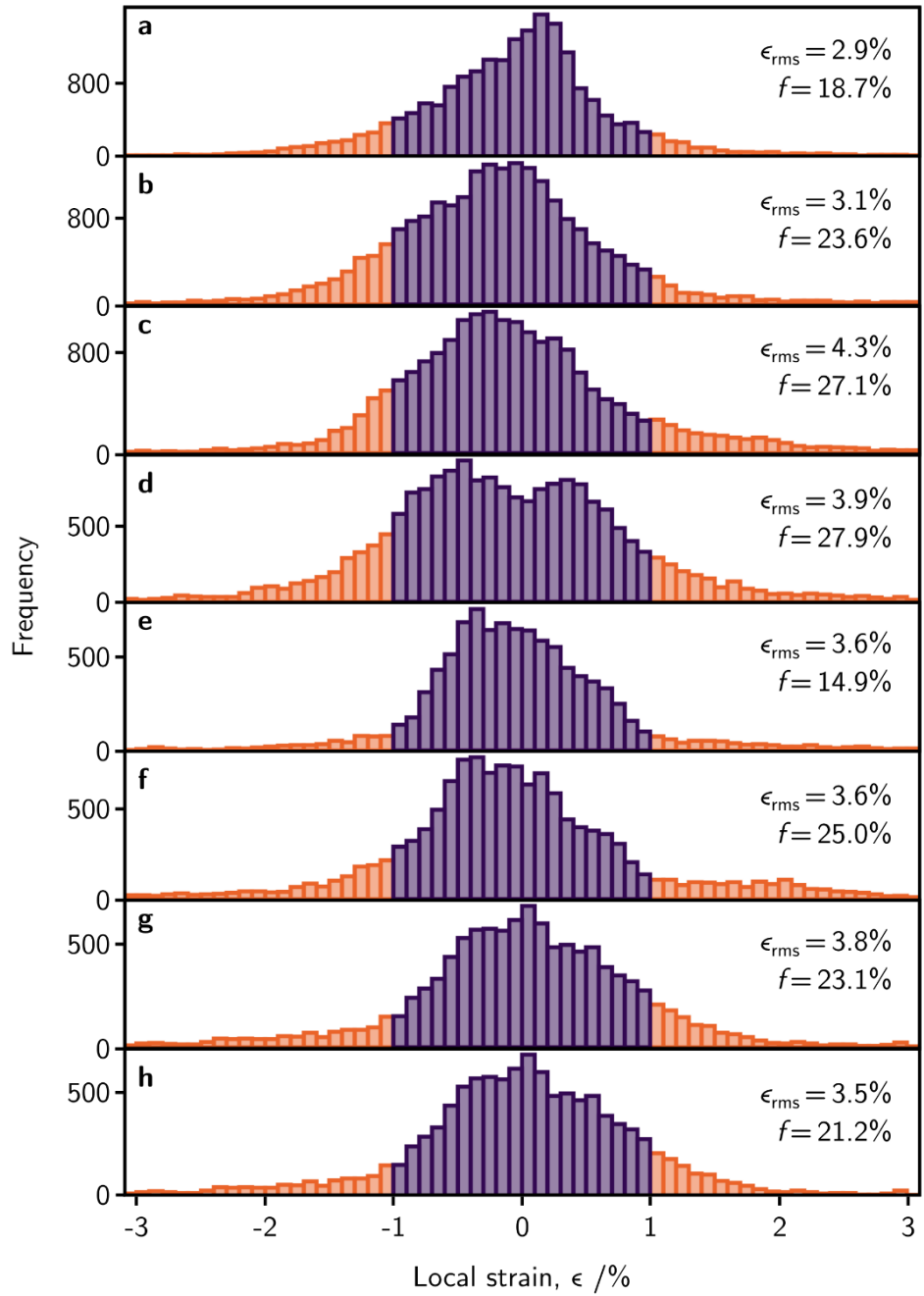

**Supplementary Fig. 25: Local strain distributions for reconstructions of the illuminated crystal:** Histograms of the local strain distribution in reconstructions shown in **a** Main text Fig. 2a & Supplementary Fig. 2a, **b** Supplementary Fig. 2b, **c** Supplementary Fig. 2c, **d** Main text Fig. 3a & Supplementary Fig. 2d, **e** Main text Fig. 3b, **f** Main text Fig. 3c, **g** Main text Fig. 3d, **h** Main text Fig. 3e. Values with a magnitude greater than 1% are highlighted in orange.

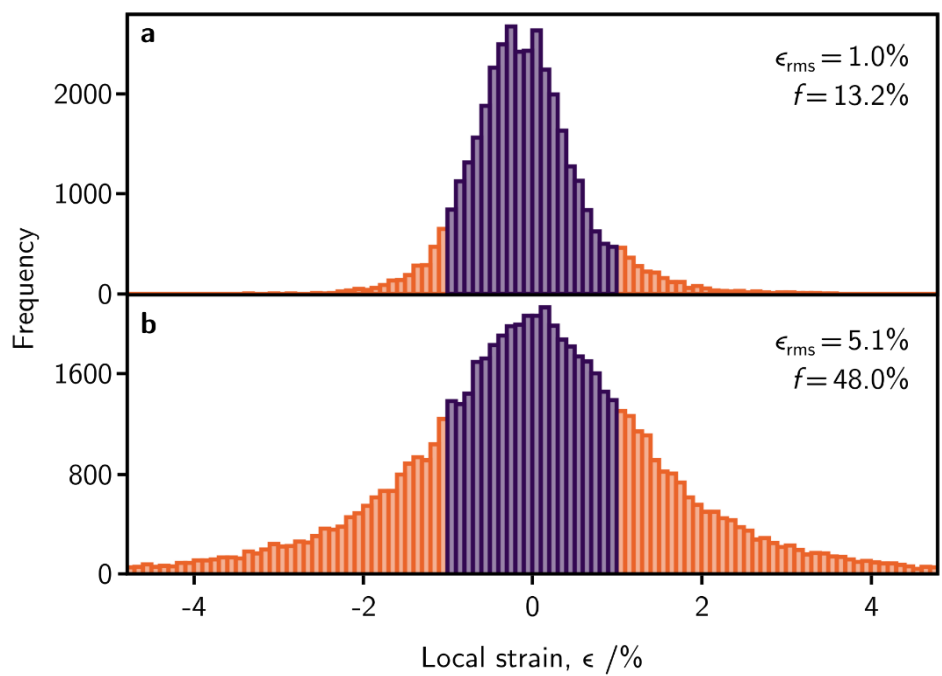

**Supplementary Fig. 26: Local strain distributions for reconstructions of the crystal shown in Fig. 4a-d of the main text:** Histograms of the local strain distribution in reconstructions shown in **a** Main text Fig. 4a & 4b, **b** Main text Fig. 4c & 4d. Values with a magnitude greater than 1% are highlighted in orange.

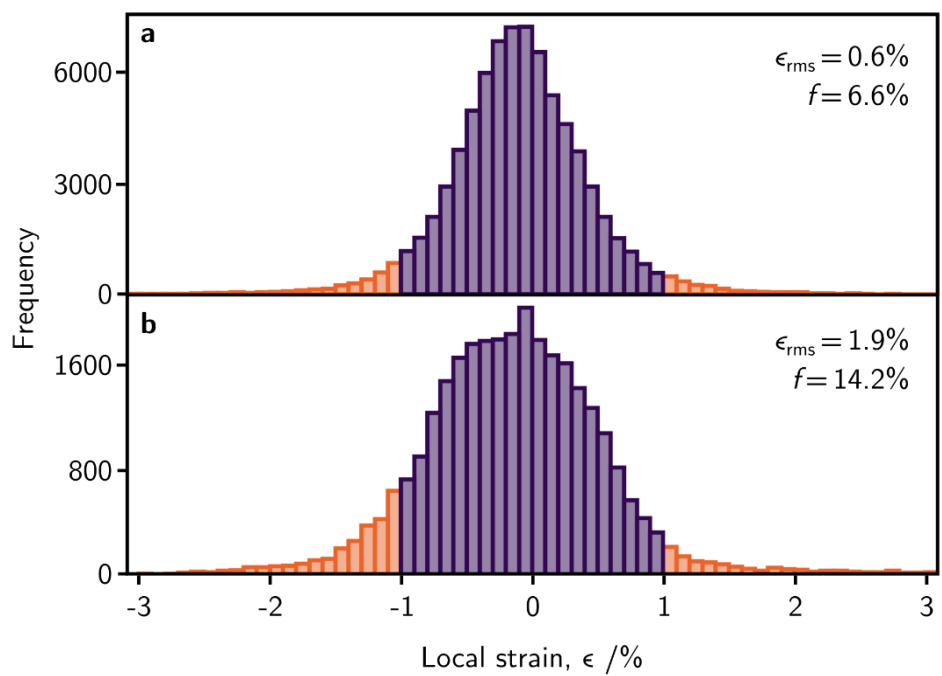

**Supplementary Fig. 27: Local strain distributions for reconstructions of the crystal shown in Supplementary Fig. 23a-d:** Histograms of the local strain distribution in reconstructions shown in **a** Supplementary Fig. 23a & 23b, **b** Supplementary Fig. 23c & 23d. Values with a magnitude greater than 1% are highlighted in orange.